# Identifying a descriptor for *d*-orbital delocalization in cathodes of Li batteries based on x-ray Compton scattering

B. Barbiellini,[1,a)] K. Suzuki,[2)] Y. Orikasa,[3)] S. Kaprzyk,[1,4)] M. Itou,[5)] K. Yamamoto,[3)] Yung Jui Wang,[1)] H. Hafiz,[1)] R. Yamada,[2)] Y. Uchimoto,[3)] A. Bansil,[1)] Y. Sakurai,[5)] and H. Sakurai[2)]

[1] *Department of Physics, Northeastern University, Boston, Massachusetts 02115, USA*

[2] *Faculty of Science and Technology, Gunma University, Kiryu, Gunma 376-8515, Japan*

[3]*Guraduate School of Human and Environmental Studies, Kyoto University, Sakyo-ku, Kyoto 606-8501, Japan*

[4]*Faculty of Physics and Applied Computer Science, AGH University of Science and Technology, aleja Mickiewicza 30, Krakow 30-059, Poland*

[5]*Japan Synchrotron Radiation Research Institute (JASRI), SPring-8, Sayo, Hyogo 679-5198, Japan*

We discuss how x-ray Compton scattering spectra can be used for investigating the evolution of electronic states in cathode materials of Li batteries under the lithiation/delithiation process. In particular, our analysis of the Compton spectra taken from polycrystalline $Li_xCoO_2$ samples shows that the spectra are dominated by the contribution of the O-2*p* redox orbital. We identify a distinct signature of *d*-orbital delocalization, which is tied directly to the conductivity of the material, providing a descriptor based on Compton spectra for monitoring the lithiation range with improved conductivity and kinetics for electrochemical operation. Our study demonstrates that Compton scattering spectroscopy can provide a window for probing complex electronic mechanisms underlying the charging and discharging processes in Li-battery materials.

____________________________

a) Electronic mail: B.Amidei@neu.edu

Scattering high energy x-rays from doped copper-oxide superconductors has provided a spectroscopic tool for unraveling correlated electronic wave functions and for directly imaging effects of impurity atoms in complex materials[1]. The technique has been adapted for investigating cathode materials in lithium batteries,[2,3] wherein also the octahedron formed by the transition metal and oxygen atoms plays a key role in controlling the electronic structure. Such studies have enabled a fundamental characterization of lithium battery materials via a combination of spectroscopy and first-principles calculations, opening a pathway for understanding the evolution of electronic states in battery materials. A challenge that needs to be addressed in this connection is that most oxide materials generally show poor electronic conductivity at ambient temperatures, which limits their performance; interestingly, however, oxides can be used as bi-functional electrocatalysts for oxygen evolution and oxygen reduction reactions[4] where also conductivity plays a crucial role. Here we address the question of how a descriptor associated with the delocalization of the *d*-electron wave function in the battery materials, which controls conductivity, can be accessed through the Compton scattering spectra.

Compton scattering refers to deep inelastic scattering between electrons and photons in the x-ray energy range. Within the impulse approximation,[5,6] the Compton scattering cross section is given by

$$\frac{d^2\sigma}{d\Omega dE_2} = F \cdot J(p_z) \quad , \tag{1}$$

where $J(p_z)$ is the Compton profile and $p_z$ is the scattering vector, which is taken to lie along the z-direction. The explicit form of the function $F$ is given by Ribberfors.[7] The Compton profile can be related to the ground-state electron momentum density $\rho(\mathbf{p})$ by the double integral

$$J(p_z) = \iint \rho(\mathbf{p}) dp_x dp_y \quad , \tag{2}$$

where $\mathbf{p} = (p_x, p_y, p_z)$ is the electron momentum. The momentum density can be expressed as[8,9]

$$\rho(\mathbf{p}) = \sum_j n_j \left| \int \Psi_j(\mathbf{r}) \exp(-i\mathbf{p} \cdot \mathbf{r}) dr \right|^2 \quad , \tag{3}$$

where $\Psi_j(\mathbf{r})$ is the wave function of the electron in state *j* and $n_j$ is the corresponding occupation number. The index *j* runs over all orbitals. The Compton technique has been widely employed for determining bulk electronic structures and Fermi surfaces of materials.[10-23] Here we discuss the application of the technique to polycrystalline Li$_x$CoO$_2$ in order to gain insight into the nature of redox orbitals associated with Li intercalation. We also propose a descriptor based on Compton profile differences to monitor *d*-orbital delocalization in real space.

Polycrystalline Li$_x$CoO$_2$ samples ($x$ = 0, 0.5, 0.625, 0.75, and 1) were prepared by extracting lithium chemically. Purchased LiCoO$_2$ was mixed with NO$_2$BF$_4$ and C$_2$H$_3$N in an Ar atmosphere glove box. After mixing, solutions were filtered and washed with C$_2$H$_3$N and dried in vacuum, and composition was determined by inductively coupled plasma (ICP) measurements. Pellets of samples were then produced via cold isostatic press. Pellets used for Compton profile measurements were of 10 mm diameter and 2 mm thickness.

The Compton scattering experiments were performed on BL08W beamline at SPring-8 using a Cauchois-type x-ray spectrometer.[24-26] The incident x-ray energy was 115 keV and the scattering angle was 165 degrees. Size of the incident x-ray beam was 1.8 mm in height and 2.0 mm in width. Measurements were performed under vacuum conditions at room temperature. The overall momentum resolution was 0.1 atomic unit (a.u.).

Density Functional Theory (DFT) based Korringa-Kohn-Rostoker coherent-potential-approximation (KKR-CPA) calculations were performed within the framework of the local spin-density-approximation (LSDA)[27,28] in order to obtain the spherically-averaged, theoretical Compton profiles of Li$_x$CoO$_2$ for various *x*-values. The effect of disorder in the Li sublattice was thus modeled via the KKR-CPA by randomly occupying Li sites by Li atoms or by leaving them vacant with probabilities reflecting the overall Li concentration.[29,30] The ensemble-averaged Green's function and its momentum matrix elements were then computed rigorously within the framework of the KKR-CPA scheme. A high degree of convergence was achieved in the KKR-CPA as well as the charge self-consistency cycles. An advantage of the KKR-CPA method is that calculations can be performed within the relatively small rhombohedral unit cell of LiCoO$_2$ with four basis atoms without the need for invoking large supercells. The system is described by spin-glass-like behavior with randomly oriented cobalt effective spin moments, while the moments on the oxygen sites are negligible. The muffin-tin radii used for Li, Co and O are 2.4, 2.6 and 1.2 a.u, respectively. The LSDA potential produces an energy gap of only about 1 eV at $x$ = 1, but a larger energy gap can be obtained by using hybrid functionals.[31] However, the Compton profiles, which are our main interest, are generally not too sensitive to the presence of band gaps in the spectrum.[1] Therefore, energy gap corrections have not been included in the KKR-CPA based theoretical profiles.

In order to carry out Mulliken population studies, we also performed DFT calculations using the CRYSTAL14 code.[31,32] This code is based on a periodic self-consistent field (SCF) approach in which Bloch wave functions are expressed in terms of localized basis-sets of atomic orbitals. The geometry used in the computations on $Li_xCoO_2$ ($x$ = 0.25, 0.33, 0.5, 0.75, and 1) was that given in Ref. 33. The exchange-correlation energy was calculated within the framework of the hybrid functional B3LYP. Gaussian basis sets were used for the Li, Co and O atoms.[34-36]

Our KKR-CPA calculations show that Li ions are strongly screened by the local oxygen environment since the Li muffin-tin sphere with a radius 2.4 a.u. contains 3.4 electrons when $x$ = 0 and 3.2 electrons for $x$ = 1. In fact, Van der Ven *et al.*[37] have shown that although the Li ions are ionized, the electron transfer from Li to the host is quite local. These authors also note a substantial increase in the electron density at the O sites immediately surrounding the Li ions; the electron density at the O sites has a distribution that resembles that of atomic $p$ orbitals, while the local charge transfer from Li to its neighboring O ions forms an almost spherical electron cloud that locally screens the positive charge of the ionized Li ion. The Mulliken analysis given in the Supplementary Material[38] clearly shows that the redox orbital associated with Li intercalation has mostly O-2$p$ character in agreement with the soft x-ray absorption spectroscopy (XAS) results of Mizokawa *et al.*[39]

Suzuki *et al.*[2] have pointed out that the changes (increase in electrons at momenta with $p_z$ < 1 a.u.) in the experimental Compton profile with Li intercalation in the spinel $Li_xMn_2O_4$ can be explained mostly in terms of the O-2$p$ redox orbitals.[2] A similar increase of electrons with low momenta is seen in the difference valence Compton profiles of $Li_xCoO_2$ in Figure 1. [By taking differences of Compton profiles, we are able to zoom in on changes in the electron occupancy near the Fermi level associated with Li intercalation/deintercalation.] Suzuki *et al.*[2] also noticed that a negative excursion between 1 and 4 a.u. in the difference Compton profile of $Li_xMn_2O_4$ can be accounted for if some $d$ electrons of Mn are transferred from localized to less localized $d$ states. Comparing results for $x$ = 1.079 and 0.496, they found that the number of electrons in the negative part of the difference Compton profile is 0.16 per lithium atom, indicating that if one electron goes from the valence of lithium to a delocalized O-2$p$ orbital, this transfer induces a wave-function modification of about 0.16 electrons in the manganese 3$d$ shell. However, for $x$ > 1, the negative part of the Compton profile difference was found to disappear, so that the behavior of the Mn-3$d$ wave functions no longer exhibits the preceding delocalization pattern.

Since the region of normal operation of the spinel $Li_xMn_2O_4$ cathode is limited over the range $x$ = 0 to 1, could the aforementioned negative excursion have a correlation with improved conductivity and kinetics for electrochemical operation?

To answer this question, here we consider the case of LiCoO$_2$, which is a layered oxide used in many commercial Li battery cathodes.[40,41] Although the material undergoes crystallographic transitions upon the removal of Li, the structure remains layered and one can obtain single-phase polycrystalline samples of CoO$_2$ and Li$_x$CoO$_2$ through electrochemical de-intercalation of Li from pure LiCoO$_2$. At $x = 0.92$, a metal-insulator transition appears. Although the samples exhibit a biphasic metal-insulator state over $x = 0.94$ to $x = 0.75$, for $x > 0.55$, ordered Li-vacancy structures can emerge.[42,43] Interestingly, reversible charge–discharge cycling of LiCoO$_2$ is only possible down to 50 % of the available Li-ions since vacancy ordering can drastically reduce the battery capacity and cycle stability. Intuitively, one expects that the Co $d$ orbitals are more localized in the insulating phase and in the vacancy ordered structures, while over the region $x = 0.6 - 0.75$, delocalization effects occur.

The measurements and the computational results given in Fig. 1 show that the negative excursion between 1 to 4 a.u. in the difference Compton profile of LiCoO$_2$ appears only when $x > 0.65$. In this regard the experimental results are in good agreement with the corresponding first-principles KKR-CPA calculations, confirming that the negative excursions occur mostly near the metallic regions. The finer features between 0 and 1.3 a.u. in the spectra of Fig. 1 are related to Fermi surface breaks in the underlying momentum density, which are smeared due to disorder,[44,45] electron correlation and experimental resolution effects.[8,9]

Figure 2 shows a fitting analysis of the Compton profile differences (upper figures) and the corresponding kinetic energy profiles $E_{KE}(p)$ (lower figures). This analysis was performed using two curves: an atomic O-2$p$ Compton profile[46] and a Compton profile, which reflects 3$d$ orbital delocalization in real space at the Co sites. The radial wave functions used for the analysis are Slater-type orbitals detailed in the Supplementary Materials.[38] The resulting modulation of the Co contribution to the profiles thus corresponds to $d$ electrons, which are transferred from localized to less localized $d$ states as shown by Suzuki et al.[2] [More accurately, the 3$d$ orbital delocalization curves in Fig. 2 reflect a redistribution of the electron momentum density where some weight is transferred from high to low momenta.] Good fits seen in Fig. 2 indicate both the important role of the O-2$p$ redox orbital[39] and that the negative excursion in the difference Compton profile reflects delocalization of the Co $d$ orbital. Moreover, the number of electrons involved in the $d$-orbital delocalization process can be obtained from the area under the negative excursion, which also equals the number of Li electrons, which are displaced from high to low momentum region. Table I shows that this number is negligible in the Compton profile differences, $\Delta J(p_z)$, between $x = 0.5$ and $x = 0.0$ as well as between $x = 1.0$ and $x = 0.75$, while the largest number is obtained for $\Delta J(p_z)$ between $x = 0.75$ and $x = 0.625$, which corresponds to the Li concentration region with improved conductivity and kinetics for electrochemical operation.

Spherically averaged Compton profiles can be used to calculate the kinetic energy. As pointed out by Epstein,[47] a formal connection between the expectation value of the electron kinetic energy and the isotropic Compton profile exists, and variations in kinetic energy can be obtained from difference profiles $\Delta J(p)$ as follows:

$$E_{\mathrm{KE}}(p) = 3\int_0^p q^2 \Delta J(q)\,dq \quad . \tag{4}$$

In the lower part of Fig. 2, the kinetic energy variation $E_{\mathrm{KE}}(p)$ produced by the Compton profile difference $\Delta J(p)$ is shown over the momentum interval of 0 to 4 a.u. $E_{\mathrm{KE}}(p)$ is seen to converge very slowly to its infinite momentum value. [Accurate studies of the kinetic energy variation for this reason are difficult.] Nevertheless, we see that the positive kinetic energy given by the O-2$p$ orbital can be compensated and sometimes even overcome by the negative kinetic energy contribution due to $d$ electron wave function delocalization. Interestingly, the strongest net kinetic energy lowering in Fig. 2 is seen in the difference [$(x = 0.75) - (x = 0.625)$]. In this way, the negative part of the Compton profile difference is also connected to the overall kinetic energy lowering.

The delocalization of Co-3$d$ states involves changes in O-2$p$ states through their hybridization with the Co-3$d$ states, which should be accessible via soft x-ray spectroscopy. Notably, as illustrated by the analysis of Fig. 2, Co-O bonding effects tend to cancel out in spherically averaged Compton profiles[5], which are thus not sensitive to different geometries related to Li intercalation, although the valence DOS exhibits small changes with varying O position (see Figure 9 in Milewska et al.[48]).

In conclusion, our study demonstrates that x-ray Compton scattering can provide a probe of the electronic structure of Li battery cathodes. In particular, our analysis identifies a descriptor for the oxide electronic conductivity in terms of the negative excursion of the (lithiated) difference Compton profiles, which monitors delocalization of the wave-function of 3$d$ Co states. This descriptor is shown to correctly capture the Li concentration region corresponding to improved conductivity and kinetics for electrochemical operation in both $LiMn_2O_4$ and $LiCoO_2$ cathodes, and will be valuable more generally for optimizing reversible electrochemical devices since poor electronic conductivity of oxide electrodes limits device performance. Notably, we expect the present descriptor to be applicable only in cases where the O-2$p$ state plays a dominant role in the lithiation redox process $O^-/O^{2-}$ while some 3$d$ electrons of the transition metal are transferred from localized to less localized 3$d$ states.[49]

**ACKNOWLEDGMENTS**

The authors thank Mr. S. Tajima of Gunma University for technical support. The work at Northeastern University was supported by the US Department of Energy (DOE), Office of Science, Basic Energy Sciences grant number DE-FG02-07ER46352, and benefited from Northeastern University's Advanced Scientific Computation Center (ASCC), and the NERSC supercomputing center through DOE grant number DE-AC02-05CH11231. The work at Gunma University, JASRI, and Kyoto University was supported by Japan Science and Technology Agency. K.S. was supported by Grant-in-Aid for Young Scientists (B) from MEXT KAKENHI under Grant Nos. 24750065 and 15K17873. The Compton scattering experiments were performed with the approval of JASRI [Proposal No. 2013B1187].

## FIGURES

Figure 1.

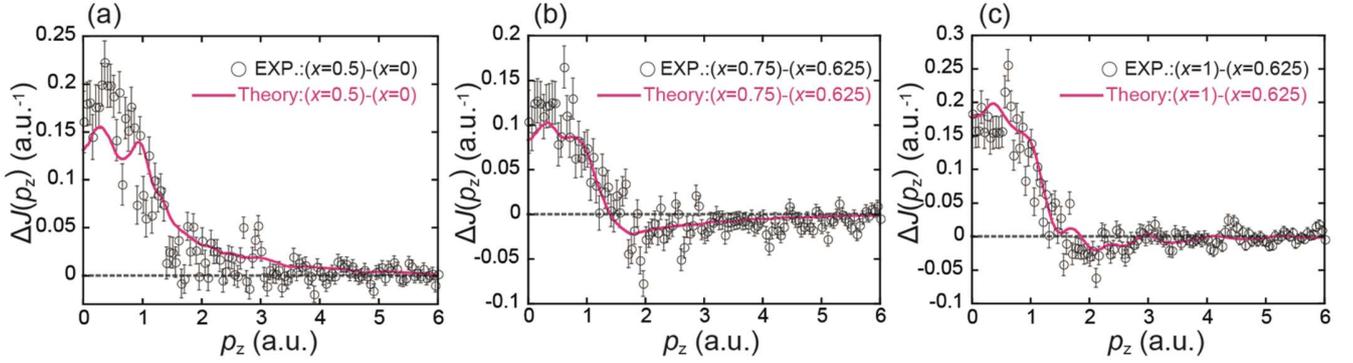

FIG. 1. (Color Online) Comparison of experimental differences of valence Compton profiles of $Li_xCoO_2$ (open circles) with the corresponding theoretical difference profiles obtained using the KKR-CPA scheme (red solid lines) for various pairs of Li concentrations $x$, as indicated in panels (a)-(c).

Figure 2.

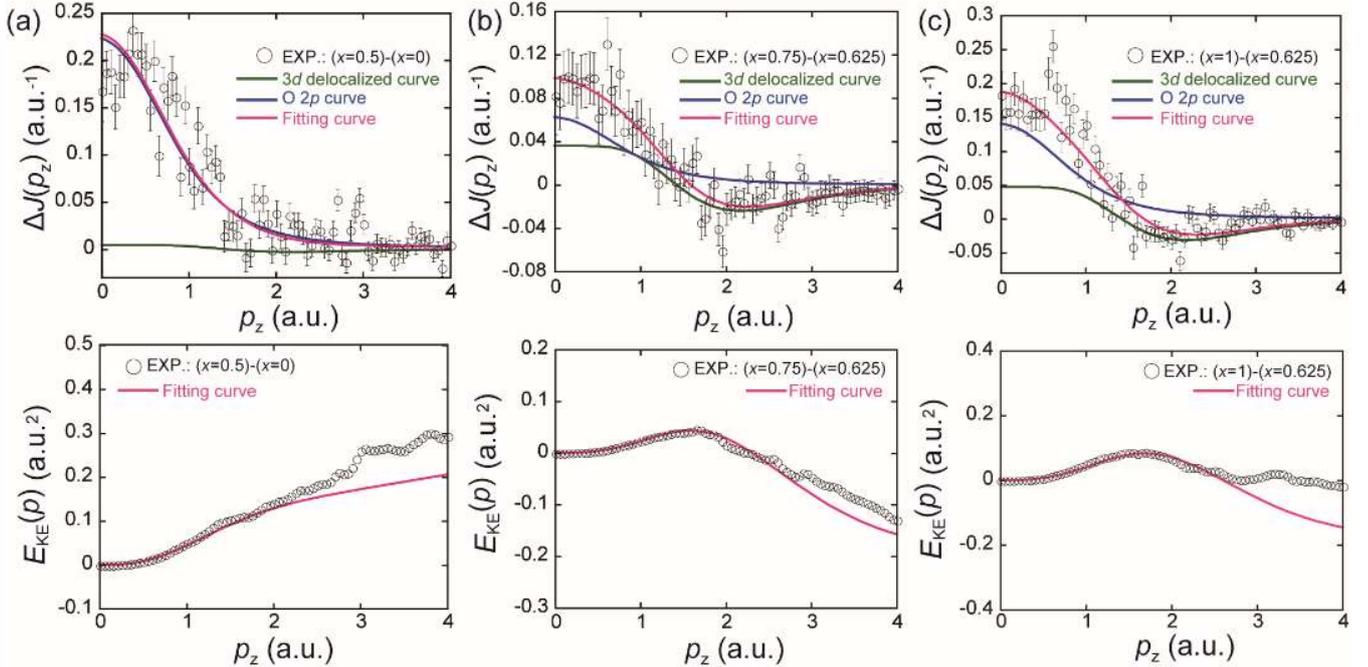

FIG. 2. (Color online) Curve fitting analysis of the Compton profile differences (upper figures) and the kinetic energy variations estimated from these differences (lower figures). Compton profile differences are composed of the O-2p redox contributions (blue solid lines) and the momentum density modulation resulting from Co-3d orbital delocalization (green solid lines) as shown in upper row of figures. The corresponding kinetic energies changes, $E_{KE}(p)$, are deduced from the fitting curves (red solid lines) and compared with the experimental data as shown in the lower row of figures.

**TABLES**

TABLE I. Number of electrons associated with the negative part of the Compton profile difference per lithium atom.

| Li concentration pair | Numbers of electrons |
|:---:|:---:|
| 0.5 – 0 | 0.021 |
| 0.75 – 0.625 | 0.609 |
| 1 – 0.625 | 0.355 |
| 1 – 0.75 | 0.028 |